\documentclass[showpacs,amssymb,aps,preprint]{revtex4}
\usepackage{amsmath}
\usepackage{amstext}
\usepackage{amsopn}
\usepackage{amsfonts}
\usepackage{amssymb}
\usepackage{bbm}
\usepackage{accents}
\usepackage{empheq}%for overbracket
\usepackage{graphicx}
\usepackage{epsf}
\usepackage{graphics}
\usepackage[latin1]{inputenc}
\def\sc{\scriptscriptstyle}

\begin{document}

\title{The structure of supersymmetry in ${\cal PT}$ symmetric quantum mechanics}

\author{D. Bazeia$^{a}$, Ashok Das$^{b,c}$, L. Greenwood$^{b}$ and L. Losano$^{a}$\footnote{e-mail: bazeia@fisica.ufpb.br, das@pas.rochester.edu,  lgreenwo@pas.rochester.edu, losano@fisica.ufpb.br}}
\affiliation{$^{a}$ Departamento de F\'{\i}sica, Universidade Federal da Para\'{\i}ba, 58051-970 Jo\~ao Pessoa, PB, Brazil}
\affiliation{$^b$ Department of Physics and Astronomy, University of Rochester, Rochester, NY 14627-0171, USA}
\affiliation{$^c$ Saha Institute of Nuclear Physics, 1/AF Bidhannagar, Calcutta 700064, India}

\begin{abstract}
The structure of supersymmetry is analyzed systematically in ${\cal PT}$ symmetric quantum mechanical theories. We give a detailed description of supersymmetric systems associated with one dimensional ${\cal PT}$ symmetric quantum mechanical theories. We  show that there is a richer structure present in these theories compared to the conventional theories associated with Hermitian Hamiltonians. We bring out various properties associated with these supersymmetric systems and generalize such quantum mechanical theories to higher dimensions as well as to the case of one dimensional shape invariant potentials.
\end{abstract}


\maketitle
\newpage

${\cal PT}$ (combined parity and time reflection) symmetric theories provide an interesting class of theories described by non-Hermitian Hamiltonians which possess real eigenvalues \cite{bender,mostafazadeh}. In the past several years  various aspects of such theories have been investigated in detail. One such interesting topic that has been explored to some extent is how to incorporate supersymmetry into one dimensional ${\cal PT}$ symmetric quantum mechanical theories \cite{susy}. However, there is yet to be a systematic discussion of this with an analysis of  the structure of supersymmetry as well as a generalization of this to quantum mechanical theories in higher dimensions. This is the purpose of this letter.

Let us recall that in one dimension, the basic transformation under ${\cal PT}$ is given by
\begin{equation}
p\stackrel{\cal\sc PT}{\longrightarrow} {\cal PT} p {\cal PT} = p,\quad ix\stackrel{\cal\sc PT}{\longrightarrow} {\cal PT} (ix) {\cal PT} = ix.\label{1}
\end{equation}
Therefore, a general one dimensional ${\cal PT}$ symmetric quantum mechanical theory is described by a Hamiltonian of the form
\begin{equation}
H = H (p, ix) = H^{\cal\sc PT} (p, ix).\label{2}
\end{equation}
In particular, we are interested in Hamiltonians of the standard form (we consider only such Hamiltonians in this paper)
\begin{equation}
H = \frac{1}{2} \left( p^{2} + V (ix)\right),\label{3}
\end{equation}
where the potential is ${\cal PT}$ symmetric, but in general may not be Hermitian. As a result, the ${\cal PT}$ symmetric Hamiltonian \eqref{3} is in general not Hermitian. However, given a ${\cal PT}$ symmetric potential $V(ix)$, there are three other Hamiltonians that one can associate with the ${\cal PT}$ symmetric Hamiltonian of \eqref{3}, namely,
\begin{eqnarray}
H^{\dagger} (p, ix) & = & \frac{1}{2}\left(p^{2} + V (-ix)\right) = {\cal P} H (p,ix) {\cal P} = {\cal T}H(p,ix){\cal T},\nonumber\\
H_{1} (p, ix) & = & \frac{1}{2}\left(p^{2} + \frac{1}{2} (V (ix) + V (-ix))\right),\nonumber\\
H_{2} (p,ix) & = & \frac{1}{2}\left(p^{2} + \frac{i}{2}(V(ix) - V (-ix))\right).\label{4}
\end{eqnarray}
Here $H^{\dagger}$ and $H_{1}$ are ${\cal PT}$ symmetric, but $H_{2}$ is not (because of the factor ``$i$" in front of the potential). (In this paper, for simplicity we use a ``dagger" to denote the formal adjoint of an operator in the Dirac sense. Since we do not use any hermiticity property, the Hamiltonian $H^{\dagger}$ in \eqref{4} can equivalently be labelled as say, $H^{\sc {\cal P}}$, to signify that it is distinct from $H$ and corresponds to its parity transform.) On the other hand, $H_{1}$ and $H_{2}$ are Hermitian while $H$ and $H^{\dagger}$ are not in general. We note that $H$ and $H^{\dagger}$ share the same eigenvalues and  that $H_{1}$ is, in fact, invariant under  ${\cal P}$ and ${\cal T}$ individually. When $V(ix)$ is in addition Hermitian, $H, H^{\dagger}$ and $H_{1}$ coincide and are ${\cal PT}$ symmetric as well as Hermitian while $H_{2}$ becomes the trivial Hamiltonian for the free particle which is both Hermitian and ${\cal PT}$ symmetric. Since we are interested in only ${\cal PT}$ symmetric theories, we will disregard $H_{2}$ from now on. Nonetheless, it is worth recognizing  that a given ${\cal PT}$ symmetric potential leads to three distinct (but related) ${\cal PT}$ symmetric Hamiltonians.

It is well known that in one dimension, a Hermitian quantum mechanical Hamiltonian with a bounded potential (from below) factorizes (up to a finite additive constant) as \cite{sturmliouville}
\begin{equation}
h = h^{\dagger} = a^{\dagger} a,\quad a = \frac{1}{\sqrt{2}}\left(p - i W (x)\right),\label{5}
\end{equation}
where the factors (and the superpotential $W(x)$) may be obtained from a knowledge of the ground state of the system. As an example, we note that the Hamiltonian for the harmonic oscillator (with an additive constant) factorizes as 
\begin{equation}
h = \frac{1}{2} \left(p^{2} + x^{2} - 1\right) = \frac{1}{2} (p + ix) (p - ix).\label{6}
\end{equation}
However, in the case of a general ${\cal PT}$ symmetric system, the Hamiltonian is not Hermitian. As a result, in one dimension a general ${\cal PT}$ symmetric Hamiltonian (with a finite additive constant) can at the best be factorized as 
\begin{equation}
H = H^{\cal\sc PT} = A^{\dagger}B \neq H^{\dagger},\label{7}
\end{equation}
where each of the factors $A$ and $B$ can be either ${\cal PT}$ symmetric or anti-symmetric. (This is already evident in the factorization in \eqref{6} and we note that if we consider factors that are linear in $p$ as in \eqref{5}, each of the factors must be ${\cal PT}$ symetric.) In general, $A$ and $B$ need not be related. However, if the ${\cal PT}$ symmetric theory is, in addition, also Hermitian, then we have $B=A$. As an example, we note that  a non-Hermitian  ${\cal PT}$ symmetric theory with a quartic term in the potential \cite{mikhail} factorizes as
\begin{equation}
H = \frac{1}{2} \left(p^{2} - x^{4} - 2ix\right) = \frac{1}{2} \left(p + (ix)^{2}\right)\left(p - (ix)^{2}\right),\label{8}
\end{equation}
where each of the factors is ${\cal PT}$ symmetric, but otherwise they are independent. 

These examples suggest that, in general, for a ${\cal PT}$ symmetric theory we can define a ${\cal PT}$ symmetric superpotential of the form
\begin{equation}
W = W (ix),\quad W^{\dagger}(ix) = W (-ix) = {\cal P} W(ix){\cal P} = {\cal T} W(ix){\cal T},\label{9}
\end{equation}
and identify
\begin{equation}
A = \frac{1}{\sqrt{2}}\left(p + W^{\dagger}(ix)\right) = \frac{1}{\sqrt{2}}\left(p+W (-ix)\right),\quad B= \frac{1}{\sqrt{2}}\left(p - W (ix)\right),\label{10}
\end{equation}
so that the factorization in \eqref{7} can also be written as
\begin{equation}
H = A^{\dagger}B = \frac{1}{2}\left(p + W (ix)\right)\left(p - W (ix)\right).\label{11}
\end{equation}
Therefore, for a general ${\cal PT}$ symmetric Hamiltonian of the form in \eqref{3} (with an additive constant), the potential $V (ix)$ is related to the superpotential $W (ix)$ through the ${\cal PT}$ symmetric Riccati relation
\begin{equation}
V (ix) = - W^{2} (ix) + i \frac{\mathrm{d} W (ix)}{\mathrm{d}x} = - W^{2} (ix) - W^{\prime} (ix),\label{12}
\end{equation}
where a ``prime" denotes taking derivative with respect to the argument $ix$ of the superpotential. We also note here that the operators $A$ and $B$ are parity conjugates of each other (up to a sign) in the sense that
\begin{equation}
A = -{\cal P} B {\cal P},\quad A^{\dagger} = - {\cal P} B^{\dagger} {\cal P}.\label{12a}
\end{equation}

It is worth emphasizing here that if a ${\cal PT}$ symmetric Hamiltonian is related to a Hermitian Hamiltonian through a similarity transformation \cite{mostafazadeh,jones}
\begin{equation}
H = S h S^{-1},\label{similarity}
\end{equation}
and if the Hermitian Hamiltonian factorizes as
\begin{equation}
h = a^{\dagger} a,
\end{equation}
then, we can identify
\begin{equation}
B = S a S^{-1},\quad A = (S^{\dagger})^{-1} a S^{\dagger}.\label{similarity1}
\end{equation}
In this case, if the similarity transformation $S$ (or $S^{-1}$) does not take a state out of the Hilbert space, then the Hamiltonian for the ${\cal PT}$ symmetric theory would share the same spectrum with the Hermitian Hamiltonian as well as inherit various of its nice features. (One particular feature that would not be inherited is the fact that the eigenstates of the ${\cal PT}$ symmetric Hamiltonian would not be orthonormal, with respect to the standard (Dirac) inner product, even though orthonormality holds for the eigenstates of the Hermitian Hamiltonian. As a result, the inner product for the ${\cal PT}$ symmetric theory needs to be redefined carefully \cite{bender,mostafazadeh}.) On the other hand, if the similarity transformation $S$ (or $S^{-1}$) takes a state out of the Hilbert space, then the nice features of the Hermitian Hamiltonian would not, in general, be shared by the ${\cal PT}$ symmetric Hamiltonian. 

Given a ${\cal PT}$ symmetric Hamiltonian described by a superpotential $W (ix)$, supersymmetric systems can be defined in a natural way \cite{witten}. We note here that in the case of a Hermitian Hamiltonian, a superpotential leads only to a pair of supersymmetric Hamiltonians (unless the potential is also shape invariant \cite{shapeinvariant}). As we will show now, the structure of supersymmetry is much richer in the case of ${\cal PT}$ symmetric theories. Let us define the supersymmetric charges
\begin{equation}
Q = A \psi,\quad \tilde{Q} = B \psi,\quad Q^{\dagger} = A^{\dagger} \psi^{\dagger},\quad \tilde{Q}^{\dagger} = B^{\dagger} \psi^{\dagger},\label{13}
\end{equation}
where $A,B$ are defined in \eqref{10} and the fermionic variables $\psi,\psi^{\dagger}$ satisfy the Clifford algebra
\begin{equation}
[\psi, \psi^{\dagger}]_{+} = 1,\quad \psi^{2} = 0 = (\psi^{\dagger})^{2}.\label{14}
\end{equation}
A particularly convenient realization for the fermionic variables is given by the Pauli matrices
\begin{equation}
\psi = \sigma_{-} = \frac{1}{2}(\sigma_{1}-i\sigma_{2}),\quad \psi^{\dagger} = \sigma_{+} = \frac{1}{2} (\sigma_{1}+i\sigma_{2}).\label{15}
\end{equation}
It is clear that by virtue of the Grassmann nilpotency properties of \eqref{14}, the supersymmetric charges are nilpotent, namely,
\begin{equation}
Q^{2} = 0 = \tilde{Q}^{2},\quad (Q^{\dagger})^{2} = 0 = (\tilde{Q}^{\dagger})^{2},\label{16}
\end{equation}
as they are expected to be. It is also straightforward to check now that the closed supersymmetry algebra
\begin{equation}
\tilde{Q}^{2} = 0 = (Q^{\dagger})^{2},\quad [Q^{\dagger}, \tilde{Q}]_{+} = H_{\sc{\rm SUSY}},\label{17}
\end{equation}
leads to the supersymmetric Hamiltonian (we do not need to write explicitly that $\tilde{Q}$ and $Q^{\dagger}$ commute with $H_{\sc{\rm SUSY}}$ which follows from the nilpotency of the supercharges)
\begin{equation}
H_{\sc{\rm SUSY}} = \begin{pmatrix}
A^{\dagger}B & 0\\
0 & BA^{\dagger}
\end{pmatrix} = \begin{pmatrix}
H_{-} & 0\\
0 & H_{+}
\end{pmatrix},\label{18}
\end{equation}
where we have identified $H_{-}$ with the Hamiltonian $H$ in \eqref{3}.

We note that $H_{\sc{\rm SUSY}}$ is invariant under supersymmetry transformations generated by $\tilde{Q}$ and $Q^{\dagger}$ (which follows from the nilpotency of the supercharges) and $H_{-}$ and $H_{+}$ denote supersymmetric partner Hamiltonians corresponding to our original Hamiltonian in \eqref{3}. It is easily seen that they are almost isospectral. Namely, if $|\psi_{n}\rangle$ denotes an eigenstate of $H_{-}$ with nonvanishing eigenvalue,
\begin{equation}
H_{-}|\psi_{n}\rangle = A^{\dagger} B |\psi_{n}\rangle = \lambda_{n} |\psi_{n}\rangle,\label{19}
\end{equation}
then, $B|\psi_{n}\rangle$ denotes the eigenstate of $H_{+}$ with the same eigenvalue,
\begin{equation}
H_{+}\left(B|\psi_{n}\rangle\right) = BA^{\dagger}B|\psi_{n}\rangle = BH_{-}|\psi_{n}\rangle = \lambda_{n} \left(B|\psi_{n}\rangle\right).\label{20}
\end{equation}
The pairing does not hold only for the state with zero energy value which is determined from
\begin{equation}
B|\psi_{0}\rangle = 0,\quad{\rm or,}\quad A^{\dagger}|\psi_{0}\rangle = 0.\label{21}
\end{equation}
Consideration of the first relation in \eqref{21}, for example, leads to the equation satisfied by the wave function of the zero eigenvalue state  of the form
\begin{equation}
\left(-i \frac{\mathrm{d}}{\mathrm{d}x} + W(ix)\right)\psi_{0} (x) = 0,\label{22}
\end{equation}
whose solution can be written as
\begin{equation}
\psi_{0} (x) \simeq e^{-i \int^{x} \mathrm{d} x'\,W(ix')} = e^{-i\int^{x} \mathrm{d}x'\left({\rm Re}\, W(ix') + i {\rm Im}\, W (ix')\right)}.\label{23}
\end{equation}
Thus, the existence of such a normalizable state does not impose any condition on the real part of the superpotential. However, the imaginary part of the superpotential must be an odd function of $x$ (with appropriate sign for damping).

It is worth recalling some of the properties of supersymmetric quantum mechanics in systems described by conventional Hermitian Hamiltonians. In such a case, the analog of the supersymmetry algebra \eqref{16} is given by
\begin{equation}
Q^{2} = 0 = (Q^{\dagger})^{2},\quad [Q,Q^{\dagger}]_{+} = H_{\sc{\rm SUSY}},
\end{equation}
and the supersymmetric  Hamiltonian is easily seen to be a positive semi-definite operator. Consequently, all the energy levels in such a case are known to be positive semi-definite with the ground state corresponding to zero energy eigenvalue. (We recall here that it is this feature that makes it quite difficult to break supersymmetry in conventional Hermitian supersymmetric theories.) On the other hand, the supersymmetry algebra for a ${\cal PT}$ symmetric theory given in \eqref{17} involves two distinct supersymmetric charges (the supercharges are not Hermitian conjugate of each other and the Hamiltonian is not a sum of positive semi-definite operators as in standard supersymmetry) and, therefore, the energy eigenvalues in such a case are not positive semi-definite in general (they can become negative). This can already be seen explicitly in the negative energy eigenvalues obtained in some of the models constructed in \cite{susy}. However, if $H_{\sc{\rm SUSY}}$ for a ${\cal PT}$ symmetric system is related to that of a Hermitian system through a similarity transformation that does not take states out of the Hilbert space, the energy eigenvalues for such a system will be positive semi-definite.

Since there are two supersymmetric charges in a ${\cal PT}$ symmetric theory, we can construct a second supersymmetric Hamiltonian from the supersymmetry algebra
\begin{equation}
Q^{2} = 0 = (\tilde{Q}^{\dagger})^{2},\quad [Q, \tilde{Q}^{\dagger}]_{+} = H_{\sc{\rm SUSY}}^{\dagger},
\end{equation}
and it is straightforward to work out explicitly the form of the Hamiltonian
\begin{equation}
H_{\sc{\rm SUSY}}^{\dagger} = \begin{pmatrix}
B^{\dagger}A & 0\\
0 & AB^{\dagger}
\end{pmatrix} = \begin{pmatrix}
H_{-}^{\dagger} & 0\\
0 & H_{+}^{\dagger}
\end{pmatrix}.
\end{equation}
We recognize this to be the Hermitian conjugate of the Hamiltonian in \eqref{18} and we note that  $H_{-}^{\dagger}$ and $H_{+}^{\dagger}$ are superpartner Hamiltonians corresponding to the supersymmetrization of $H^{\dagger}$ in \eqref{4}. As before, we note that $H_{-}^{\dagger}$ and $H_{+}^{\dagger}$ are almost isospectral. In fact, being Hermitian conjugates of each other, $H_{\sc{\rm SUSY}}$ and $H_{\sc{\rm SUSY}}^{\dagger}$ are isospectral. Namely, both $H_{-}$ and $H_{-}^{\dagger}$ have exactly the same spectrum (including zero modes if any) just as $H_{+}$ and $H_{+}^{\dagger}$ do. In fact,  it is easy to check that if $|\psi\rangle$ denotes an eigenstate of $H_{-}$, namely,
\begin{equation}
H_{-} |\psi\rangle = A^{\dagger}B|\psi\rangle = \lambda |\psi\rangle.
\end{equation}
then, using \eqref{12a} we obtain 
\begin{equation}
H_{-}^{\dagger}({\cal P}|\psi\rangle) = B^{\dagger}A{\cal P}|\psi\rangle = {\cal P} A^{\dagger}B|\psi\rangle = \lambda({\cal P}|\psi\rangle).
\end{equation}
Similarly, the eigenstates of $H_{+}$ and $H_{+}^{\dagger}$ are related by the parity operator ${\cal P}$. (It is worth emphasizing here that ${\cal P}$ does not take a state out of the Hilbert space.) 

It would seem natural that the Hermitian Hamiltonian $H_{1}$ in \eqref{4} can also be supersymmetrized using the relation
\begin{equation}
\frac{1}{2}[\tilde{Q},Q^{\dagger}]_{+} + \frac{1}{2} [Q, \tilde{Q}^{\dagger}]_{+} = H_{1\,{\sc{\rm SUSY}}},
\end{equation}
with the supersymmetric Hamiltonian of the form
\begin{equation}
H_{1\,{\sc{\rm SUSY}}} = \frac{1}{2}\begin{pmatrix}
A^{\dagger}B + B^{\dagger}A & 0\\
0 & AB^{\dagger} + BA^{\dagger}\end{pmatrix} = \begin{pmatrix}
H_{1\, -} & 0\\
0 & H_{1\, +}
\end{pmatrix}.
\end{equation}
However, the supersymmetry of the system is not easy to see in this basis. The reason for this is quite simple. As we have noted earlier, the Hamiltonian $H_{1}$ is invariant under ${\cal P}$ and ${\cal T}$ independently and the supercharges should reflect this as well. Therefore, let us define the supercharges
\begin{equation}
\overline{Q} = \bar{A} \sigma_{-} = \frac{1}{\sqrt{2}} \left(p - \overline{W} (ix)\right)\sigma_{-},\quad \overline{Q}^{\dagger} = \bar{A}^{\dagger} \sigma_{+} = \frac{1}{\sqrt{2}} \left(p - \overline{W} (-ix)\right)\sigma_{+},\label{A}
\end{equation}
where $\overline{W} (-ix) = - \overline{W} (ix)$ is related to the superpotential $W (ix)$ through the relation
\begin{equation}
i \frac{\mathrm{d}\overline{W}(ix)}{\mathrm{d}x} - \overline{W}^{2} (ix)  = \frac{1}{2}\left[i \frac{\mathrm{d} (W (ix) - W (-ix))}{\mathrm{d} x} - (W^{2} (ix) + W^{2} (-ix))\right].
\end{equation}
It is clear that each of the factors $\bar{A}$ and $\bar{A}^{\dagger}$ in \eqref{A} changes sign (anti-symmetric) under ${\cal P}$ and ${\cal T}$ independently and is ${\cal PT}$ symmetric. We can now define the supersymmetry algebra
\begin{equation}
\overline{Q}^{2} = 0 = (\overline{Q}^{\dagger})^{2},\quad [\overline{Q},\overline{Q}^{\dagger}]_{+} = H_{1\,{\sc\rm SUSY}},
\end{equation}
with the explicit form for the supersymmetric Hamiltonian
\begin{equation}
H_{1\,{\sc\rm SUSY}} = \begin{pmatrix}
H_{1\, -} & 0\\
0 & H_{1\, +}
\end{pmatrix} = \begin{pmatrix}
\bar{A}^{\dagger} \bar{A} & 0\\
0 & \bar{A} \bar{A}^{\dagger}
\end{pmatrix}.
\end{equation}
The isospectrality of the supersymmetric partner Hamiltonians is now obvious and the energy eigenvalues in this case would be positive semi-definite.

It is worth pointing out that the same superpotential $W (ix)$ (and the supercharges $Q, \tilde{Q}$) allow us to construct other supersymmetric Hamiltonians as well. For example, the superalgebra
\begin{equation}
\tilde{Q}^{2} = 0 = (\tilde{Q}^{\dagger})^{2},\quad [\tilde{Q},\tilde{Q}^{\dagger}]_{+} = H_{\sc B\,{\rm SUSY}},
\end{equation}
leads to the supersymmetric Hamiltonian of the form
\begin{equation}
H_{\sc B\,{\rm SUSY}} = \begin{pmatrix}
B^{\dagger} B & 0\\
0 & B B^{\dagger}
\end{pmatrix} = \begin{pmatrix}
H_{{\sc B}\, -} & 0\\
0 & H_{{\sc B}\, +}
\end{pmatrix}.\label{28}
\end{equation}
Similarly, the supersymmetry algebra
\begin{equation}
Q^{2} = 0 = (Q^{\dagger})^{2},\quad [Q, Q^{\dagger}]_{+} = H_{\sc A\,{\rm SUSY}},
\end{equation}
where the supersymmetric Hamiltonian has the form
\begin{equation}
H_{\sc A\,{\rm SUSY}} = \begin{pmatrix}
A^{\dagger} A & 0\\
0 & A A^{\dagger}
\end{pmatrix} = \begin{pmatrix}
H_{{\sc A}\, -} & 0\\
0 & H_{{\sc A}\, +}
\end{pmatrix}.
\end{equation}
Both $H_{\sc B\,{\rm SUSY}}$ and $H_{\sc A\,{\rm SUSY}}$ are Hermitian and, therefore, would have positive semi-definite energy eigenvalues. Furthermore, we note  from \eqref{12a} that
\begin{equation}
H_{\sc A\,{\rm SUSY}} = {\cal P} H_{\sc B\,{\rm SUSY}} {\cal P},
\end{equation}
so that it is sufficient to study the structure of only one of the Hamiltonians, for example $H_{\sc B\,{\rm SUSY}}$ in \eqref{28}. Using the definition of $B$ in \eqref{10} we can show that
\begin{eqnarray}
H_{{\sc B}\,-} &=& B^{\dagger}B =  \frac{1}{2}\left((p - {\rm Re}\, W (ix))^{2} + ({\rm Im}\, W(ix))^{2} - i {\rm Im}\, W^{\prime} (ix)\right),\nonumber\\
H_{{\sc B}\,{+}} &=& BB^{\dagger} =  \frac{1}{2}\left((p - {\rm Re}\, W (ix))^{2} + ({\rm Im}\, W(ix))^{2} + i {\rm Im}\, W^{\prime} (ix)\right).
\end{eqnarray}
It is clear now that these Hamiltonians behave like a particle coupled to an external Abelian gauge field with the real part of the superpotential behaving like the gauge potential \cite{das}. Thus, we see that in the case of ${\cal PT}$ symmetric theories, there is a rich structure of supersymmetric systems -- the same superpotential $W(ix)$ leads to a larger number of supersymmetric Hamiltonians than in the conventional case of a Hermitian Hamiltonian.

In going to higher dimensional ${\cal PT}$ symmetric quantum mechanical  theories with supersymmetry, we note that in two dimensions, parity can only be defined as reflecting 
\begin{equation}
x \stackrel{\sc\cal P}{\longrightarrow} -x.\quad {\rm or,}\quad y \stackrel{\sc\cal P}{\longrightarrow} - y,
\end{equation}
but not both. (It is worth recalling here the well known fact that in odd space-time dimensions, reflecting all coordinates, namely, reflecting through the origin is equivalent to a rotation. For example, for two space dimensions, $x\rightarrow -x, y\rightarrow -y$ is equivalent to a rotation around the $z$ axis by an angle $\pi$. In such a case, a discrete parity transformation representing a disconnected Lorentz transformation with determinant $-1$ can be defined by reflecting only one of the coordinates.) As a result, if we want a theory with rotational invariance, the ${\cal PT}$ symmetric Hamiltonian would coincide with conventional Hermitian theories and the discussion of supersymmetric theories does not lead to anything new. The Landau problem would be an example of a supersymmetric two dimensional ${\cal PT}$ symmetric theory. In going to quantum mechanical theories in three dimensions, we can introduce three superpotentials $W_{k} (i\mathbf{x}), k=1,2,3,$ (in this case, parity reflects all space coordinates, $\mathbf{x}\rightarrow - \mathbf{x}$) and generalize the definition of the supercharges in \eqref{13} as
\begin{equation}
Q = A_{k} \psi_{k},\quad \tilde{Q} = B_{k} \psi_{k},\quad Q^{\dagger} = A_{k}^{\dagger} \psi_{k}^{\dagger},\quad \tilde{Q}^{\dagger} = B_{k}^{\dagger} \psi_{k}^{\dagger},\label{29}
\end{equation}
where repeated indices are summed and 
\begin{equation}
A_{k} = \frac{1}{\sqrt{2}}\left(p_{k} + W_{k}^{\dagger}(i\mathbf{x})\right) = \frac{1}{\sqrt{2}}\left(p_{k}+W_{k} (-i\mathbf{x})\right),\quad B_{k}= \frac{1}{\sqrt{2}}\left(p_{k} - W_{k} (i\mathbf{x})\right).
\end{equation}
A realization for the fermion variables in this case, can be given by enlarging the matrix space as \cite{das,jorge}
\begin{equation}
\psi_{k} = \sigma_{k}\otimes \sigma_{-},\quad \psi_{k}^{\dagger} = \sigma_{k}\otimes \sigma_{+},
\end{equation}
which satisfy
\begin{equation}
\psi_{j}\psi_{k} = 0 = \psi_{j}^{\dagger}\psi_{k}^{\dagger},\quad j,k = 1,2,3,\label{30}
\end{equation}
and this is essential in the definition of a higher dimensional supersymmetry algebra. 

By virtue of the nilpotency in \eqref{30}, it now follows that
\begin{equation}
Q^{2} = \tilde{Q}^{2} = 0,\quad (Q^{\dagger})^{2} = (\tilde{Q}^{\dagger})^{2} = 0,
\end{equation}
and the construction of supersymmetric systems can be carried out as in the case of quantum mechanical theories in one dimension. We will discuss only one of the systems for completeness. For example, if we consider the supersymmetry algebra (see \eqref{17})
\begin{equation}
\tilde{Q}^{2} = 0 = (Q^{\dagger})^{2},\quad [Q^{\dagger}, \tilde{Q}]_{+} = H_{\sc\rm SUSY},
\end{equation}
the supersymmetric Hamiltonian can be explicitly calculated and in this case takes the block diagonal form
\begin{equation}
H_{\sc\rm SUSY} = \begin{pmatrix}
H_{-} & 0\\
0 & H_{+}
\end{pmatrix} = \begin{pmatrix}
A_{k}^{\dagger}B_{k}\, \mathbbm{1}_{2} + i \epsilon_{klm} A_{k}^{\dagger}B_{l}\sigma_{m} & 0\\
0 & B_{k} A_{k}^{\dagger}\, \mathbbm{1}_{2} + i \epsilon_{klm} B_{k}A_{l}^{\dagger}\sigma_{m}
\end{pmatrix},
\end{equation}
where $k,l,m = 1,2,3$ and repeated indices are summed. Other supersymmetric systems can similarly be constructed.

Supersymmetry naturally leads to the discussion of shape invariance. For example, let us consider the one dimensional supersymmetric system described in \eqref{17} and \eqref{18} and denote
\begin{eqnarray}
H_{-} & = & A^{\dagger} B = \frac{1}{2}\left[p^{2} + V_{-} (ix)\right] = \frac{1}{2}\left[p^{2} -  W^{\prime} (ix) - W^{2} (ix)\right],\nonumber\\
H_{+} & = & B A^{\dagger} = \frac{1}{2}\left[p^{2} + V_{+} (ix)\right] = \frac{1}{2}\left[p^{2} + W^{\prime} (ix) - W^{2} (ix)\right].\label{shapeinv}
\end{eqnarray}
The potential $V_{-} (ix)$, of course, depends on some parameters such as the coupling constant which we denote by writing $V_{-} (ix, a_{0})$. If it so happens that
\begin{equation}
V_{+} (ix, a_{0}) = V_{-} (ix, a_{1}) + R(a_{1}),
\end{equation}
where $a_{1}$ is a given function of $a_{0}$, namely, $a_{1} = f(a_{0})$ and $R (a_{1})$ is a constant, then we say that the potential is shape invariant \cite{shapeinvariant}. Namely, in such a case, the supersymmetric partner Hamiltonian has  the potential with the same shape as the potential in the original Hamiltonian, with a different parameter and with possible shift (scaling).  In this case,  we note from \eqref{shapeinv} that we can write
\begin{equation}
H_{+} (a_{0}) = B (a_{0}) A^{\dagger} (a_{0}) = A^{\dagger} (a_{1}) B (a_{1}) + R (a_{1}) = H_{-} (a_{1}) + R (a_{1}).
\end{equation}
As a result, we can construct a sequence of Hamiltonians, any two consecutive ones of which being almost isospectral, namely,
\begin{eqnarray}
H^{(0)} & = & H_{-} (a_{0}),\nonumber\\
H^{(1)} & = & H_{+} (a_{0}) = H_{-} (a_{1}) + R (a_{1}),\nonumber\\
\vdots & = & \vdots\nonumber\\
H^{(s)} & = & H_{+} (a_{s-1}) + \sum\limits_{k=1}^{s-1} R (a_{k}) = H_{-} (a_{s}) + \sum\limits_{k=1}^{s} R (a_{k}),
\end{eqnarray}
where $a_{s} = f^{s} (a_{0}) = f ( f( \cdots f(a_{0})\cdots )$. Any two consecutive Hamiltonians in this sequence are almost isospectral which allows one to determine the energy levels of the original Hamiltonian $H_{-} (a_{0})$ simply as \cite{shapeinvariant,dasbook}
\begin{equation}
E_{n} = \sum\limits_{k=1}^{n} R (a_{k}).
\end{equation}
If the sequence of Hamiltonians ends in a system that can be explicitly solved, then, the original system can also be solved. The important thing to note in the case of ${\cal PT}$ symmetric theories is that if $V_{+} (ix, a_{0})$ is shape invariant, so is $V_{+}^{\dagger} (ix, a_{0})$ and so just like supersymmetric systems that we have talked about, we can also construct other shape invariant ${\cal PT}$ symmetric systems associated with the original system. 

The construction that we have described so far is very general and we would like to end this discussion  with two examples. For simplicity, both these examples would involve ${\cal PT}$ symmetric Hamiltonians that are related to Hermitian Hamiltonians through a similarity transformation (see \eqref{similarity}). First, let us consider the ${\cal PT}$ symmetric Hamiltonian
\begin{equation}
H = \frac{1}{2}\left(p^{2} + x^{2} - 2i\epsilon x - (1+\epsilon^{2})\right) = S\, \frac{1}{2}\left(p^{2} + x^{2} -1\right) S^{-1} = S h S^{-1},\label{example1}
\end{equation}
where $\epsilon$ is a real constant and the similarity transformation is given by
\begin{equation}
S = S^{\dagger} = e^{\epsilon p}, S^{-1} = (S^{-1})^{\dagger} = e^{-\epsilon p}.\label{similarity2}
\end{equation}
(Note that, in this case, $S$  does not take a state out of the Hilbert space.) Denoting by $a, a^{\dagger}$ the annihilation and the creation operators for the harmonic oscillator (see \eqref{6}),
\begin{equation}
a = \frac{1}{\sqrt{2}}\left(p -ix\right),\quad a^{\dagger} = \frac{1}{\sqrt{2}}\left(p + ix\right),
\end{equation}
we determine (see \eqref{similarity1})
\begin{equation}
B = e^{\epsilon p} a e^{-\epsilon p} = \frac{1}{\sqrt{2}}\left(p - ix - \epsilon\right),\quad A = e^{-\epsilon p} a e^{\epsilon p} = \frac{1}{\sqrt{2}}\left(p - ix + \epsilon\right).
\end{equation}
The supersymmetric partner Hamiltonians in \eqref{18}, in this case, follow to be (this is a trivially shape invariant system)
\begin{equation}
H_{-} = A^{\dagger} B = e^{\epsilon p} a^{\dagger} a\,e^{-\epsilon p},\quad H_{+} = B A^{\dagger} = e^{\epsilon p} a a^{\dagger}\,e^{-\epsilon p} = H_{-} + 1.
\end{equation}
It follows that the energy eigenvalues and (normalized) eigenstates of the superpartner Hamiltonians are given by
\begin{eqnarray}
|\psi_{n}^{-}\rangle & = & e^{\epsilon p}|n\rangle = e^{\frac{\epsilon}{\sqrt{2}} (a + a^{\dagger})} |n\rangle,\;\;E_{n\,-} = n,\nonumber\\
|\psi_{n}^{+}\rangle & = & \sqrt{n+1}\, e^{\epsilon p} |n\rangle = \sqrt{n+1}\, e^{\frac{\epsilon}{\sqrt{2}}(a+a^{\dagger})} |n\rangle,\;\;E_{n\,+} = n + 1,\quad n=0,1,2,\cdots,\label{coherent}
\end{eqnarray}
where $|n\rangle$ denotes the number eigenstates of the harmonic oscillator Hamiltonian. This construction makes clear that the supersymmetric Hamiltonian for this ${\cal PT}$ symmetric system inherits the nice features of the supersymmetric oscillator Hamiltonian and the energy eigenvalues are positive semi-definite. However, since the energy eigenstates in \eqref{coherent} are simply coherent states, they cease to be orthonormal with respect to the standard (Dirac) inner product (of the Hermitian theory).

As a second example, let us consider the ${\cal PT}$ symmetric Hamiltonian
\begin{eqnarray}
H & = & \frac{1}{2}\left(p^{2} - (ix + \epsilon)^{2} - \frac{a_{0}(a_{0}+1)}{(ix+\epsilon)^{2}} + 2a_{0} -1\right)\nonumber\\
& = & S\, \frac{1}{2} \left(p^{2} + x^{2} + \frac{a_{0}(a_{0}+1)}{x^{2}} + 2a_{0} -1\right) S^{-1},\label{example2}
\end{eqnarray}
where the similarity transformation $S$ is given in \eqref{similarity2} (it is defined within the Hilbert space). In this case, the supersymmetric partners can be constructed as before and have the forms (in this case, $W(ix) = ix+\epsilon - \frac{a_{0}}{(ix+\epsilon)}$)
\begin{eqnarray}
H_{-}(a_{0}) & = & A^{\dagger} B = \frac{1}{2} \left(p^{2} - (ix + \epsilon)^{2} - \frac{a_{0}(a_{0}+1)}{(ix+\epsilon)^{2}} + 2a_{0} -1\right),\nonumber\\
H_{+} (a_{0}) & = & B A^{\dagger} = \frac{1}{2} \left(p^{2} - (ix + \epsilon)^{2} - \frac{a_{0}(a_{0}-1)}{(ix+\epsilon)^{2}} + 2a_{0} +1\right)\nonumber\\
& = & \frac{1}{2} \left(p^{2} - (ix + \epsilon)^{2} - \frac{a_{1}(a_{1}+1)}{(ix+\epsilon)^{2}} + 2a_{1} -1\right) + R(a_{1}) = H_{-} (a_{1}) + R (a_{1}),
\end{eqnarray}
where we can identify 
\begin{equation}
a_{1} = a_{0} -1,\quad R(a_{1}) = 2.
\end{equation}
In this case, we see that the potential is shape invariant and as discussed earlier, we can construct a sequence of partner Hamiltonians any two consecutive ones of which are almost isospectral. Furthermore, if $a_{0} = n$, then the $n$th Hamiltonian in the sequence would correspond to the Hamiltonian in \eqref{example1} (up to an additive constant) whose spectrum we have already determined. Therefore, the spectrum of the Hamiltonian in \eqref{example2} can now be completely determined following our general discussion.

To conclude, we have given a systematic description of supersymmetric systems associated with one dimensional ${\cal PT}$ symmetric quantum mechanical theories. We have shown that there is a richer structure present in these theories compared to the conventional supersymmetric theories associated with Hermitian Hamiltonians. We have brought out various properties associated with these supersymmetric systems and have generalized this construction to quantum mechanical theories in  higher dimensions as well as to the case of one dimensional shape invariant potentials.

\bigskip

\noindent{\bf Acknowledgments}

This work was supported in part 
by US DOE Grant number DE-FG 02-91ER40685. 
We are grateful to CAPES, CNPq, FAPESP and PRONEX-CNPq-FAPESQ, Brazil, for financial support.

%\bibliographystyle{prsty}
%\bibliography{all_new}

\end{document}